University of
**BRISTOL**

# INVESTIGATION OF HELMHOLTZ RESONATORS WITH CURVED TAPERED EMBEDDED NECK EXTENSION

Zhanlu (Louie) Chen

Department of Aerospace Engineering, University of Bristol, Queen's Building, University Walk, Bristol. BS8 1TR. UK.

## ABSTRACT

*In this study, the performance of Helmholtz resonators with curved tapered neck extensions was investigated through numerical and experimental methods. A numerical parametric analysis was carried out using the Finite Element Method with COMSOL, while experimental validation was conducted using a two-microphone acoustic impedance tube. The analysis evaluated the effects of neck outlet/inlet diameter ratio, neck length, and diameter of neck on the absorption coefficients and resonance frequency of the resonators. Results revealed that a tapered curved embedded neck configuration led to a lower absorption frequency as the outlet/inlet ratio increased trend to convergent neck geometry. Furthermore, a longer effective neck length and smaller neck diameter enhanced low-frequency noise absorption. Experimental validation with four 3D printed samples confirmed these trends.*

## 1. INTRODUCTION

Since the beginning of the jet engine age, noise generated by aircraft has emerged as a significant concern for pilots, passengers, airports, and airlines. With the development of structural, aerodynamical and material technology, the design and manufacture of aircraft has become more commercial and is intensified to focus on noise control to comply with stringent noise regulations. Therefore, the issue of noise control was raised and put challenges for aerospace industries, leads to take more research on the aircraft noise control [1][2]. In terms of current trends of aircraft industry, high bypass ratio engines are commonly used in civil industry and the fan noise become the major noise source from engine [3]. To damp the noise, acoustic liners are set up on the internal wall of the engine nacelles with series of Helmholtz resonators (HRs) which developed under the principle of Helmholtz resonance, with specific resonance frequency:

$$f_r = \frac{c_0}{2\pi} \sqrt{\frac{A_n}{V_c(l_n + \delta_n)}} \tag{1}$$

Where $f_r$ is the resonance frequency, $c_0$ is the speed of sound, $A_n$ is the cross area of the neck, $V_c$ is the volume of resonator, $l_n$ is the length of neck and $\delta_n$ is the end correlation factor. For cylindrical or rectangular necks, the end correlation calculated as 3/10 of the hydraulic diameter of the neck [4]. The Equation (1) indicated the resonance frequency was governed by the volume of cavity, the cross-sectional area and length of the resonator neck in the geometry of the resonator. Figure 1 illustrate the basic geometry of a circular concentric HRs, where $l_c$ is the height of the cylinder cavity, $l_n$ is the length of neck, $d_c$ is the diameter of the cylinder cavity, and $d_n$ is the diameter of the neck. From these, the resonance frequency of the HR can be identified with the formula (2):

$$f_r = \frac{c_0}{2\pi} \sqrt{\frac{d_n^2}{d_c^2 l_c(l_n + \delta_n)}} \tag{2}$$

The Helmholtz Resonator is well known as an effective device to reduce the noise around the resonance frequency, especially for low frequency range. This phenomenon can be explained through the principles of resonance. When sound waves enter the neck of a Helmholtz resonator, they cause the air in the neck to oscillate. This oscillation creates pressure fluctuations in the cavity, which in turn generates an opposing acoustic wave that exits the neck. At the resonance frequency, the phase difference between the incoming and opposing waves is such that they destructively interfere with each other, leading to a reduction in sound pressure levels. By this, certain frequency of noise could be reduced accordingly to the geometry of the resonator. Focus on the basic geometry, Komkin et al. [5] investigate the relationship of sound absorption correspond to the depth of cavity and neck length. Based on the basic principle, Xu et.al. investigated the habit of dual Helmholtz resonator under two degree-of-freedom system [6], and Cai et.al. did the further analysis and comparison of dual and lined Helmholtz resonator array configuration [7]. Further analysis of the performance of Helmholtz resonator with multiple necks been conducted by Langfeldt et al.[8], and explain the situation if extra leakage appeared on the resonators. Focus on the neck internal wall, Duan et al.[9] investigate the roughened neck to achieve the perfect absorption. During different stage of each flight, the power of the engine varies, and the frequency of the noise varies accordingly. To achieve the noise reduction through the whole flight circle, Gourdon [10] investigate the nonlinear neck of the resonator to achieve a boarder range of sound absorption for a single Helmholtz resonator cell. Alternatively, the rigid cell walls of the acoustic liner been investigated with replacement of flexible polymer forms to achieve boarder absorption range [11]. Focus on the cabin of the aircraft, Kone et al. [12] design a resonator with complex cavity design to achieve multi-tonal resonance frequency to reduce noise.

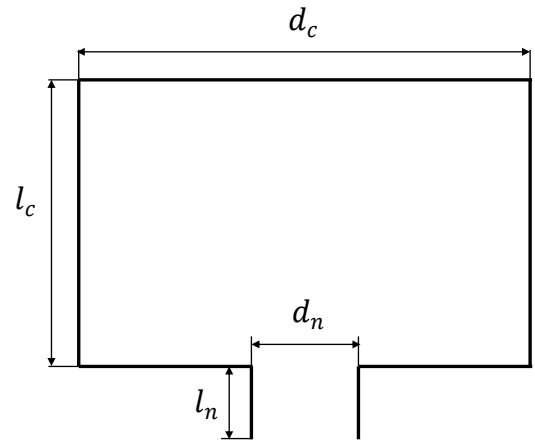

Figure 1: The basic geometry of a Helmholtz resonator.

To reduce the low-frequency noise by decrease the resonance frequency of the resonator, several studies were conducted. Selamet and Lee [13] did the experiments on extended the necks and shew the resonance frequency decrease as the neck extension increase. In their investigation, resonance frequency of concentric circular HR with straight solid, tapered, and perforated neck extension inside the cavity were compared. Shi and Mak [14] investigated the spiral neck and shew that the resonance frequency can also be reduced within a limited space. Cai et al. [15] compared the extended straight neck with spiral neck and found a spiral neck is equivalent to a straight neck with same neck length and cross-section area. This is more effectively proof that the spiral neck can be used for the HRs in limited space. Guo et al. [16] developed an acoustic metasurface with embedded spiral neck of different length, which both reduce the thickness of resonator and wider bandwidth of resonance frequency. Zhao et al. [17] investigated the effects of extended neck configurations, length, and grazing flow Mach numbers under a grazing flow for HRs. Besides from investigate the neck, the geometry of the cavity also been interested. Cambonie et al. [18] take research on bending the cavity into curve, based on this, Chen et al. [19] developed a sound absorption panel with the resonator cavity which is coiled. These supports a new angle of view to reduce the absorption frequency under keeping the panels thin.

## Problem Definition

Tang [20] suggested tapered necks have significant improvement on the sound absorption, and Song et al. [21] designed a HR with extended tapered neck inside the cavity. The effect of curved neck has been studied in past [14]– [16] and the use of longer neck to reduce resonance frequency has been noted, but there are no studies on the performance of the resonator based on a curved tapered neck extension. The aim of the research is to investigate the acoustic performance and property of HRs with tapered curved neck extensions numerically, and valid the numerical result with practical experiment. In this research, a parametric analysis take place with numerical method, and experiment with curved tapered embedded neck extension samples been conducted with impedance tube to do the validation of numerical results.

**Objectives:**

1). Numerical investigate of the effect of the tapered neck on embedded curved neck on transmission coefficient and resonance frequency.

2). Compare the acoustic performance of HRs under different configuration.

3). Validate the numerical result with experiment.

## 2. METHODOLOGY

### 2.1 Research Design and Baseline Model

The research study is aimed to investigate the performance of Helmholtz resonators (HRs) with curved tapered embedded neck extensions, focus on analysis the absorption coefficient and resonance frequency of HRs under different configurations of the embedded extended neck. To achieve the objective, a mixed-methods approach consisting of both numerical and experimental analyses been employed, with a focus on conducting a parametric study to isolate the effects of key design parameters. Figure 2 illustrate an example of the HR with tapered curved embedded neck. In this research, a parametric study been carried with three variables of the parameters been interested: the effective length of neck $l_{eff}$, the inlet diameter $d_{n1}$ of the neck and the diameter of the neck under consistent ratio $d_{n2}/d_{n1}$. All other parameters been consistent for all resonator designs, including the geometry parameter of the cavity $l_c$ and $d_c$.

The research design been divided into four main stages:

1. **Preliminary stage**: In this stage, an extensive literature review is conducted to gather relevant information on the fundamentals of Helmholtz resonators, with relevant literatures about different geometries of resonators, including the previous studies on resonators with extended neck, embedded extended neck, curved neck, and tapered necks. Further review of literatures about the numerical and experimental methods of computing the absorption coefficient and resonance frequency also been take place. This stage helps identify knowledge gaps and provides a solid foundation for the subsequent research stages.

2. **Numerical analysis stage**: A series of numerical simulations is performed using the COMSOL Multiphysics version 5.6 software with Finite Element Methods (FEM) model of both the impedance tube and HRs. A parametric study is conducted in this stage to investigate the effects of varying the effective length of neck, the inlet/outlet ratio, and the diameter of the neck, while keeping other parameters constant. The numerical analysis allows for a systematic evaluation of the effects of geometric parameters on the resonator's performance, and it facilitates the identification of the characteristic of the curved tapered neck configuration.

3. **Experimental analysis stage**: Based on the findings from the numerical analysis, selected resonator designs representing different combinations of neck diameter, neck length, and ratio of inlet/outlet diameters are 3D printed and tested using a two-microphone acoustic impedance tube, followed by the guidance of ASTM E1050-19 [22]. The experimental results are compared with the numerical predictions to validate the accuracy of the numerical model. This stage helps establish the reliability of the findings and provides insights into the practical implementation of the curved tapered embedded neck resonators in real-world applications.

4. **Data analysis and interpretation:** The results from both the numerical simulations and experimental tests are analysed to understand the individual impacts of the effective length of neck, inlet/outlet ratios of diameters of necks and the overall diameter of neck under constant ratio. Trends and relationships between these parameters and the absorption coefficient and resonance frequency are identified, providing insights for optimizing and perfect the design of Helmholtz resonators with small sizes and low resonance frequencies.

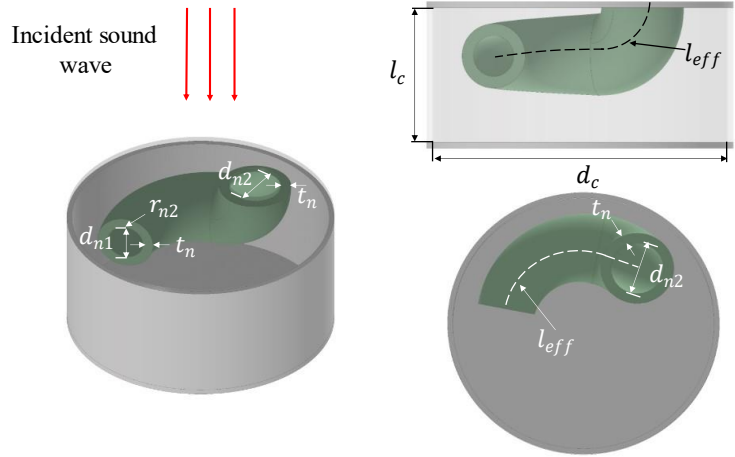

Figure 2: Illustration of the Helmholtz resonators with the tapered curved embedded neck, where $l_{eff}$ is the effective length of the neck, $d_{n1}$ is the inlet dimeter, $d_{n2}$ is the outlet dimeter, $t_n$ is the thickness of the neck tube, $l_c$ is the height of cavity and $d_c$ is the diameter of the cavity.

A baseline model of standard HR with same cavity but without the extended neck is developed and been compared with the tapered curved embedded neck configuration. From this, the effect on the reduction of resonance frequency by the embedded extended neck could be compared and evaluated. Since the baseline model will be both analysed numerically and experimentally, it should design to sit inside the measuring range of impedance tube and the geometry should fit the equipment (described in section **2.3**). Table 1 shows the parameters of the baseline model used in this analysis, with the geometry reference to the Figure 1 in introduction. By numerical analysis with COMSOL Multiphysics (section **2.2**), the baseline model follows the requirement and could been used both in numerical and experimental analysis. According to Equation (2) listed in introduction, the theoretical resonance frequency is calculated as 929.4Hz.

Table 1: The parameters of the baseline model.

| Parameters: | $l_c$ | $d_c$ | $d_n$ | $l_n$ | Theoretical $f_r$ |
|---|---|---|---|---|---|
| Value: | 23mm | 50mm | 10mm | 2mm | 929.4 Hz |

## 2.2 Numerical Analysis

### 2.2.1 Simulations Setup and Assumptions

The numerical investigation was conducted utilizing COMSOL Multiphysics, employing finite element method (FEM) models to determine transmission and absorption coefficients, subsequently identifying the resonance frequency via the graphical representation of the absorption coefficient as a function of sound wave frequency. Figure 3a describe the principle of measuring the sound wave absorption. With plane sound waves generate from left end of impedance tube and reflect backwards, the power of incident $P_{in}$ and power of outgoing $P_{out}$ at the outlet of resonator neck been computed to analyse the sound absorption. The pressure acoustic and thermos-viscous modules, illustrated in Figure 3b, facilitated the generation of plane sound waves within the impedance tube and the computation of thermal and viscous effects within the resonator respectively. For carrying the parametric analysis, displayed by Figure 3b, the height of cavity $l_c = 23mm$, diameter of cavity $d_c = 50mm$ are constant through the whole analysis. In order to establish a simulation model approximating a realistic experimental situation, the fluid in the model adopted the material properties of air, with a uniform density, sound wave speed of 343 m/s, and an absolute pressure of 1 atm. The temperature been defined as 293.15K for the model. The sound wave is sourced from the port boundary condition at far left of the impedance tube, normal to the entrance of the HR, with amplitude of 2 Pa. Another port signifying pressure release condition was applied at the outlet of the neck extension, representing an open environment where sound pressure can dissipate freely.

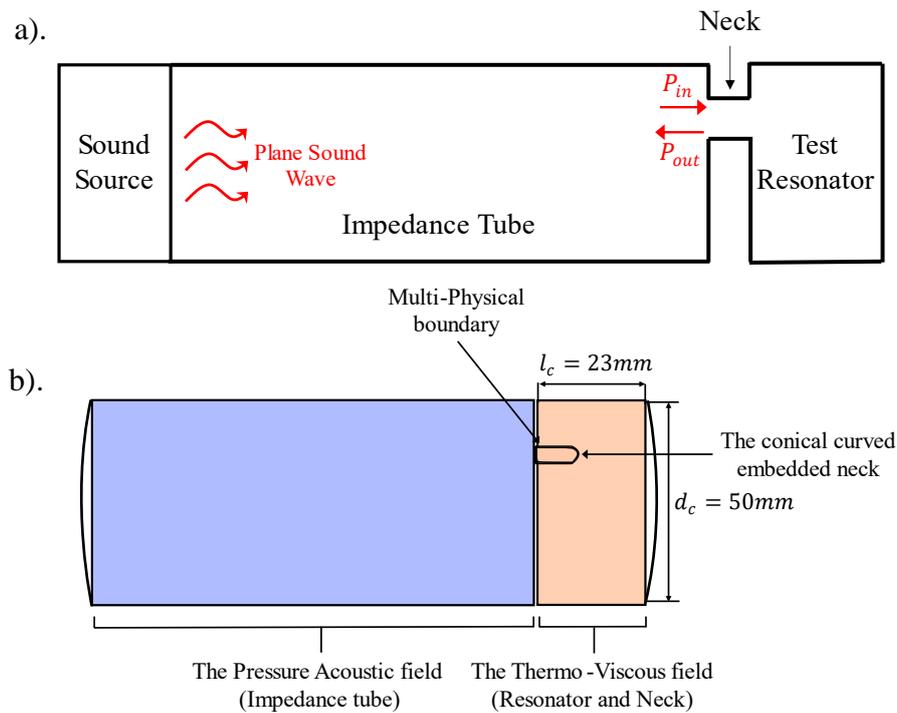

Figure 3: a). The visualised principle of sound absorption measurement. b). The illustration of side cross-sectional view of the impedance tube-resonator combination, and modules used in COMSOL Multiphysics.

For simplify the analysis and reduce the computational costs, certain assumptions been made. The fluid model been assumed as linear acoustics, meaning that the changes in pressure, density, and particle velocity are linearly related. Additionally, the fluid within the resonator is considered as isothermal, meaning that temperature variations within the resonator are negligible, and the speed of sound remains constant. Appropriate boundary conditions were applied to the model to represent the physical behaviour of the HR. Sound-hard walls were used for the impedance tube surfaces, implying that the walls are rigid and do not absorb sound energy. The resonator and neck walls were considered non-slip walls, employed to simulate boundary layers near the wall.

*2.2.2 Mesh Setup and Convergence Analysis*

To employee the Finite Element Method in COMSOL Multiphysics, The HR and impedance tube been discretized by a free tetrahedral mesh, as it is well approximate to the complex surfaces, i.e., the inside and outside wall of the tapered curved embedded neck. To ensure there are adequate number of elements inside one wavelength, to accurately resolve the variables within the sound wave, the maximum element size (22.87mm) been set, which is shorter than the 1/5 of the wavelength (343mm) of the maximum of the interested frequency range (1000Hz) [23]. The minimum element size been set at 1.14mm to ensure sufficient resolution in the complex geometric regions, e.g., the neck, without significantly increase the computational costs. The maximum element growth rate been set as 1.3, with curvature factor of 0.3 and the resolution of narrow region been set as 2. A boundary layer mesh been applied at the resonator to ensure a dense element distribution been applied at the wall of the resonator to capture more accurate boundary layer properties. Figure 4a shows the distribution of elements for the resonator and impedance tube model.

Upon determining the appropriate mesh size for this analysis, a mesh convergence study was conducted to examine the influence of mesh size on numerical outcomes. Since the COMSOL Multiphysics' mesh generation process relies on minimum and maximum mesh size, element growth rate, curvature factor, and resolution of narrow regions, mesh density cannot be easily controlled and quantified. Therefore, the chosen mesh quality been compared with fine, extra fine, and extremely fine mesh setup which have higher mesh quality than chosen. To perform the mesh sensitivity test, mesh quality was altered and the numerical results, i.e., the computed resonance frequency, were plotted for convergence analysis. If the mesh converges for the baseline model, it will also converge for other models, as they share the same acoustic properties and possess similar geometry. Figure 4b displays the resonance frequency results of the baseline resonator, exhibiting convergence at 802 Hz. The selected mesh quality, as previously described, yielded a value of 801 Hz, with an error of 0.246% compared to the convergence value (802 Hz), which is negligible and therefore confirms mesh quality validity. Considering that increased mesh quality would significantly increase the computation cost, the chosen mesh quality was deemed suitable based on the described setup. Noticed that the numerical result of the baseline model has 13.815% of error to the theoretical result obtained by Equation (2), this is due to the simplicity of the theoretical computation model, and the formula is not representable under the consideration of boundary layers and thermos-viscous effect.

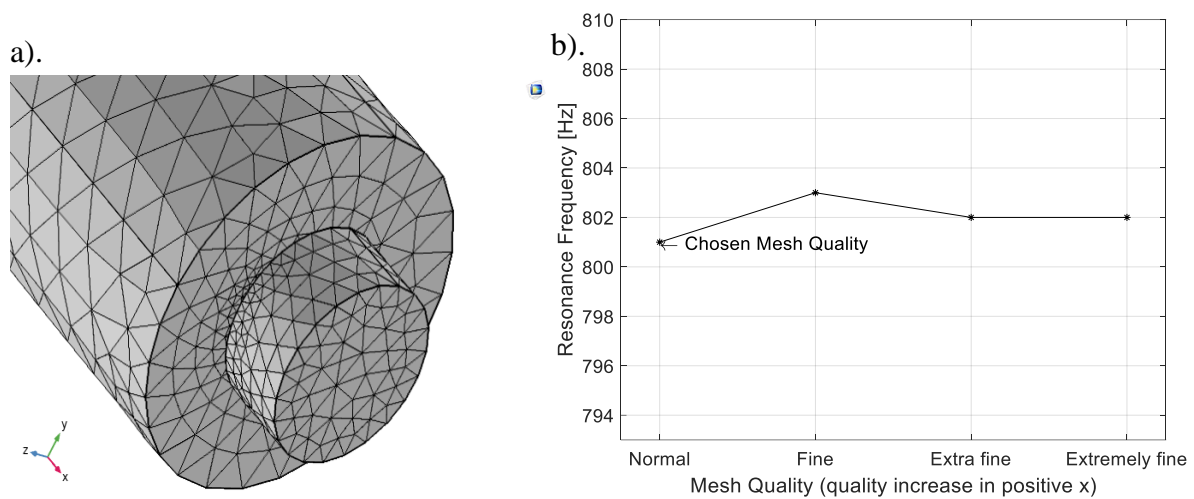

Figure 4: a). Free Tetrahedral mesh used in this COMSOL numerical analysis. b). The convergence of the resonance frequency of the baseline resonator, where mesh quality increase as x-axis increase to right.

### 2.2.3 Data Analysis

Following the completion of the numerical simulations in COMSOL Multiphysics, post-processing and data analysis techniques were employed to extract meaningful information and identify trends in the performance of Helmholtz resonators with curved tapered neck extensions. The parameter reflection coefficient $R$ been introduced first, which represent the mount of wave been reflected during the transmission of wave through the impedance discontinuity medium. The pressure acoustics module of COMSOL can directly present the power of incidence at the sound wave source, and power of outgoing at the outlet of the HR. Therefore, the reflection coefficient can be computed directly by Equation (3):

$$R = \frac{P_{out}}{P_{in}} \tag{3}$$

Where $R$ is the reflection coefficient, $P_{out}$ is the power of outgoing and $P_{in}$ is the power of incident, illustrated in Figure 3a. Therefore, the value of absorption coefficient could be computed accordingly with the reflection coefficient, following by the formula described in Equation (4):

$$\alpha = 1 - |R|^2 \tag{4}$$

Where $\alpha$ is the parameter of absorption coefficient.

The absorption coefficient is a measure of the proportion of sound energy absorbed by the resonator at a given frequency. At the resonance frequency, the resonator absorbs the maximum amount of sound energy due to the increased effectiveness of the destructive interference between the incoming and opposing waves. This results in the highest absorption coefficient at the resonance frequency. Therefore, the resonance frequency of the resonator could be obtained by plotting the graph of absorption coefficient against frequency and find the peak.

### 2.2.4 Model Validation

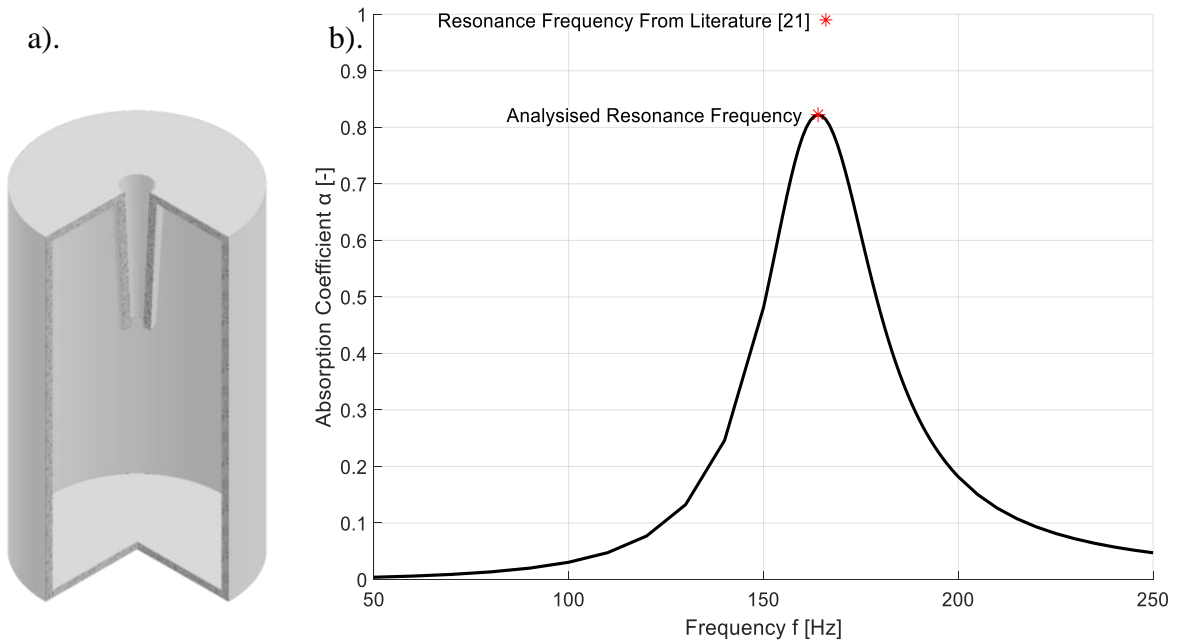

Figure 5: a). The geometry of the NTHR model from literature [21] used for validation. b). The plot of absorption coefficient for the NTHR [21] analysed by the set up COMSOL model.

The validation process of the COMSOL model is to ensure the accuracy and reliability of the numerical simulations conducted for the study. To conduct validation for the numerical methods, an existing model of a tapered neck resonator from literature [21] been analysed with the COMSOL model which setup previously, and the result been compared against the result presented by Song et.al. [21]. In this literature, an embedded neck tapered Helmholtz resonator (NTHR) been analysed with theoretical, experimental, and numerical method. The result obtained with these three methods align with high agreement, indicates that the results are very close to the real situation. Figure 5a illustrate NTHR model used for this validation, which have cylinder cavity and tapered embedded neck, similar to the models aimed to analysis in this research. Then, the model been analysed numerically through COMSOL, and plot a graph of absorption coefficient against frequency illustrated by Figure 5b, indicated that the resonance frequency located at 164Hz with absorption coefficient of 0.8224.

Table 2: Comparation between COMSOL simulation result and result from literature.

|  | Numerical Result | Result from Literature | Difference [%] |
|---|---|---|---|
| Absorption Coefficient $\alpha$ | 0.8224 | 0.99 [21] | 16.9 |
| Resonance Frequency $f$ | 164 Hz | 166 Hz [21] | 1.2 |

The referenced date from literature reported an absorption coefficient of 0.99 and a resonance frequency of 166 Hz [21]. Indicated by Figure 5b, there is a slight discrepancy between the simulation results and the reference data, the overall agreement between the two datasets is reasonably good. The resonance frequency is only 2 Hz off from the reference value, indicating that the model captures the resonant behaviour of the Helmholtz resonator quite accurately. However, the difference in the absorption coefficient warrants further investigation. Table 2 indicated that there are 16.9% of difference compared to the reference value. Potential reasons could be due to the difference of model assumptions between the COMSOL model and the reference study. Further investigation of the impact of impedance tube model set up been conducted. Surprisingly, the magnitude of absorption coefficient has slightly increase as the impedance tube diameter decrease. This phenomenon could be due to the set up of the port area in the COMSOL model is dependent by the diameter of impedance tube model, and power of outgoing is directly related to area, resulting the value change of absorption coefficient. These remind that the diameter of impedance tube should remain consist through the whole analysation. However, the literature does not present the diameter of the impedance tube model in detail, thus, the data error on absorption coefficient could not been verified in relation to the diameter of impedance tube.

In conclusion, the simulated data for resonance frequency is validated to be reliable, but the value of absorption coefficient at resonance frequency could not deliver a precise value. However, the value of absorption coefficient can still present a similar trend under this numerical analysis as the numerical model consists, and parallel comparison between different models is valid to conduct.

## 2.3 Experimental Analysis

### 2.3.1 Testing Facility

After employ the simulation, the resonators with tapered curved embedded neck extensions been investigated experimentally with University of Bristol Two-Microphone Acoustic Impedance Tube, guided by the ASTM E1050-19 [22]. This equipment can hold 100mm diameter circular sample, and able to capture the frequency range 200-2000Hz. A speaker been placed at the right end of the impedance tube to act as a sound wave source to perform plane wave inside the tube. The circular specimen been placed at the far-left side of impedance tube, and a rigid backing plate been placed at the back of specimen to create a sound-reflective termination [22]. Two microphones with gap 68mm been placed at the side wall of tube to measure the sound pressure level at same time level, and the data been collected with the National Instruments PXIe-1062Q data acquisition system. Further data analysis been conducted with MATLAB to obtain the absorption coefficient and resonance frequency. Figure 6 shows the layout of the equipment with the impedance tube.

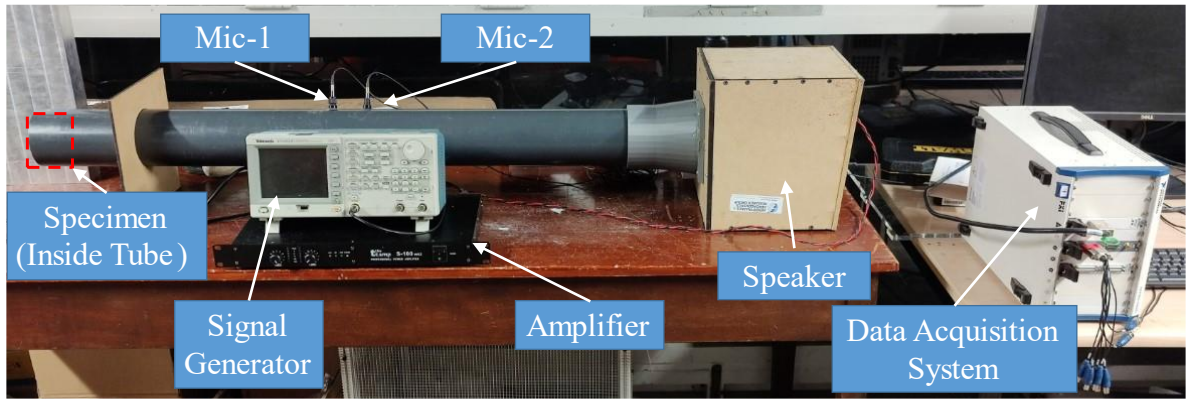

Figure 6: Layout of equipment.

### 2.3.2 Data Acquisition and Analysis

The data acquisitioned from the acquisition system been analysed through MATLAB, with the background theory developed by Chung et al. [24] and follow the procedure from ASTM E1050-19 [22]. Firstly, the speed of sound $c$ at experimental environment is obtained by formula (5), with the experiment ambient temperature $T = 20\ ℃$:

$$c = 20.047\sqrt{273.15 + T} \tag{5}$$

The MATLAB `.tfestimate` function been used to estimate the transfer functions $H$ at set of frequencies $f$ with output signal of two microphones' voltages from the data acquisition system. Figure 7 is the flow diagram which describe how the data been collected and processed.

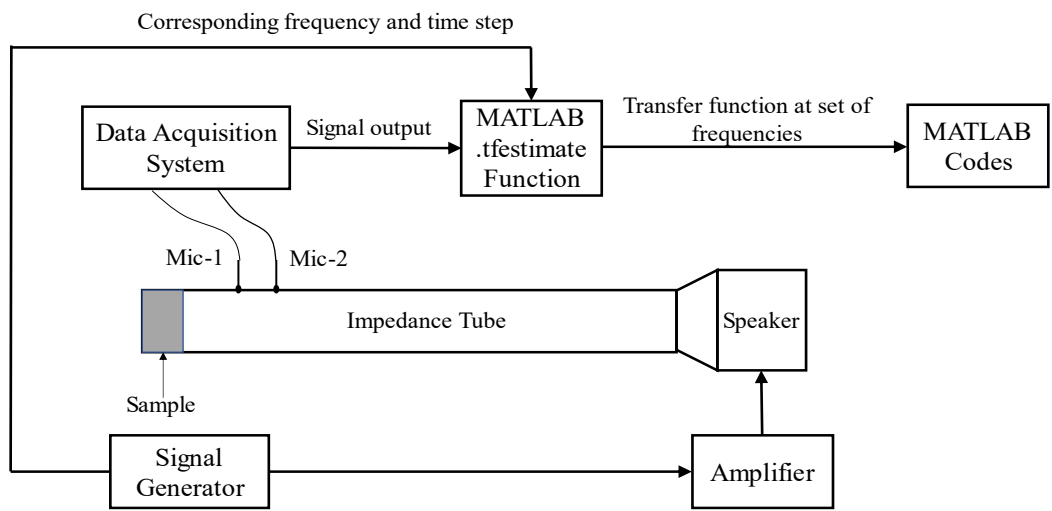

Figure 7: Flow diagram of the experiment.

For computing the reflection coefficient $R$ for the test sample, Equation (6) is used with the complex acoustic transfer function $H$:

$$R = \frac{H - e^{-j\,ks}}{e^{-j\,ks} - H} e^{-j\,2k(l+s)} \qquad (6)\ [22]$$

Where the $R$ is the complex reflection coefficient, $k$ is the wave number where $k = 2\pi f/c$, $s$ is the space between the two microphones and $l$ is the distance between the sample and the nearest microphone, where $l = 495mm$ in this experiment. Subsequently, the absorption coefficient could be obtained by Equation (7):

$$\alpha = 1 - |R|^2 \qquad (7)\ [22]$$

Where $\alpha$ is the absorption coefficient at frequency $f$. By plotting absorption coefficient against frequency, the resonance frequency could be identified which sit at the maximum of the absorption coefficient.

### 2.3.3 Test Sample Design and Manufacture

To validate the numerical analysis and investigate the performance of HR with curved tapered neck extensions in real applications, four samples (Sample 1-Sample 4) were fabricated using 3D printing technology, shown by Figure 8. The samples were designed based on the parametric study conducted in the numerical analysis, focusing on varying parameters on inlet diameter $d_{n1}$ and neck length $l_{eff}$. The model been designed by choosing the typical model from the numerical parametric analysis, with cavity height $l_c = 23mm$ and diameter $d_c = 50mm$. 3D printing technique with Polylactic Acid (PLA) material was used to produce the samples, with high-dense of layers ensure to form the hard wall boundary for the resonator. The outer geometry of the samples is a cylinder with 100mm diameter, which is exactly same as the inner size of the impedance tube. Extra lids are printed with same diameter of the sample, this is to perform the bottom plate to seal and form a complete resonator cavity geometry. The lids are glued with the body by elastic flexible superglue, which provide a tight seal to avoid leakage. Due to the accuracy of the 3D printed is not ideal, extra sanding is applied to the assembled model to fit the impedance tube.

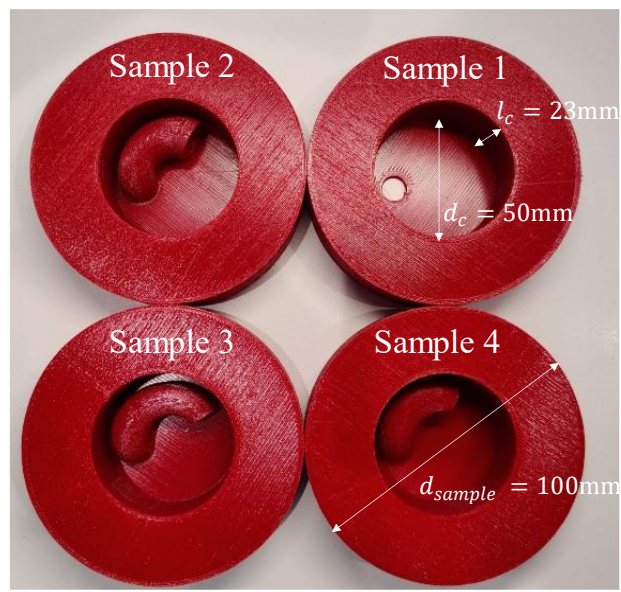

Figure 8: The photograph of the samples of the four resonators, including the baseline model.

Table 3: Data of the four resonator samples.

|  | Sample 1 | Sample 2 | Sample 3 | Sample 4 |
|---|---|---|---|---|
| $d_{n1}$ [mm] |  | 12 | 7 | 7 |
| $d_{n2}$ [mm] | Follow Baseline | 10 | 10 | 10 |
| $l_{eff}$ [mm] | Model | 36.582 | 46.060 | 36.582 |
| $d_{n2}/d_{n1}$ |  | 0.833 | 1.429 | 1.429 |
| Convergent/Divergent |  | Div | Con | Con |

# 3. RESULTS AND DISCUSSION

After further analysis of the initial data, the results of the investigation of Helmholtz resonators with curved tapered neck extensions are presented in this section. Both numerical and experimental analyses were conducted to evaluate the performance of the resonators. The primary focus was on the effects of the inlet/outlet ratio of the neck, the length of the neck, and the neck diameter on the absorption coefficients and resonance frequency of the resonators.

## 3.1 Numerical Parametric Analysis

The parametric study is conducted with the FEM numerical method (described in section 2.2) with COMSOL to investigate the performance of the resonator. The results are presented in the form of graphs showing the absorption and transmission coefficients as functions of frequency for different parameter values. In order to investigate the effect of the tapered neck compared to the normal curved neck, A design of the Helmholtz resonator with tapered curved extended neck H1 been compared with baseline model in Figure 9. The resonator H1 has effective length $l_{eff} = 36.582$ mm, with outlet and inlet diameter listed in Table 4. From observation from Figure 9, H1 has lower resonance frequencies 315Hz, and lower maximum absorption coefficients 0.9027 compared to the baseline model, which has 60.7% reduction of resonance frequency compared to baseline model with same outer cavity geometry. This suggests that the introduction of the embedded curved tapered neck shape may lead to enhanced low-frequency noise attenuation capabilities of the resonator, which could be particularly beneficial in applications where low-frequency noise reduction is critical.

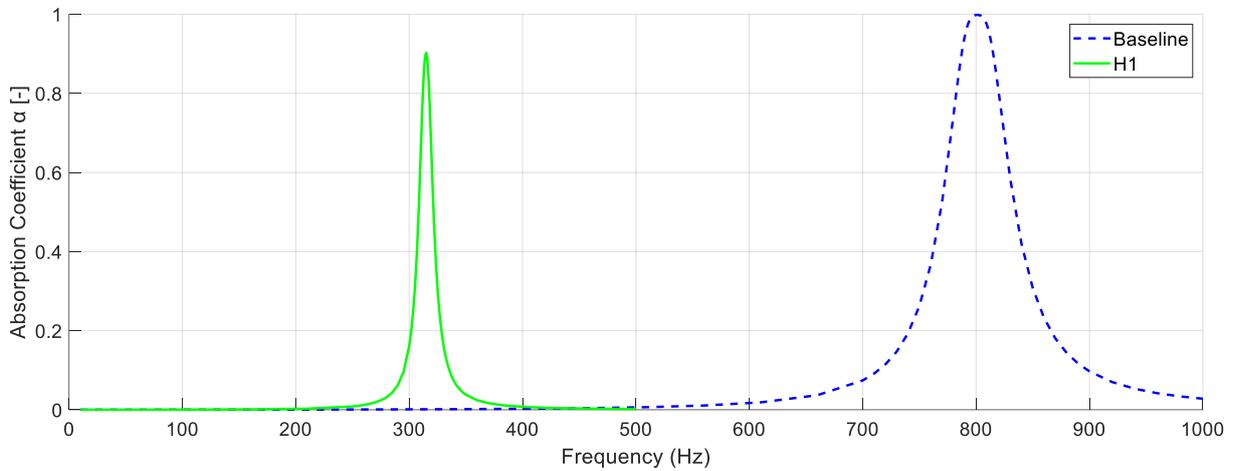

Figure 9: Comparation between tapered curved model H1 with the baseline.

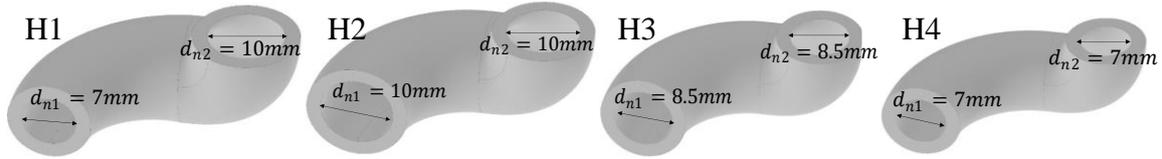

Figure 10: The extended neck geometry of the model H1-H4.

Table 4: Parameters of diameters for H1-H4.

|  | H1 | H2 | H3 | H4 |
|---|---|---|---|---|
| $d_{n1}$ [mm] | 7 | 10 | 8.5 | 7 |
| $d_{n2}$ [mm] | 10 | 10 | 8.5 | 7 |
| Tapered/non-tapered | Tapered | Non-tapered | Non-tapered | Non-tapered |

The property of the tapered neck was investigated next. The resonator H1 been compared with three non-tapered configurations H2, H3 and H4 list in Table 4. The diameter of H2 equal to the outlet diameter of H1, diameter of H4 equal to the inlet diameter of H1 and diameter of H3 is the average of inlet and outlet diameter of H1. All other properties of the resonator keep constant. Figure 11 illustrate the absorption coefficient of these four resonator models. As expected, H2, H3 and H4 have maximum value of 0.9535 at 360Hz, 0.8931 at 303Hz, 0.7519 at 248Hz respectively. The resonance frequency value of the tapered neck example sits in 315Hz between H2 and H3, and with higher absorption (0.9027) to H3. Different to the analysis done by Song et al. [21] with the straight tapered neck resonator, the resonance frequency of tapered neck resonator is slightly lower than the resonator with average diameter in this literature. This phenomenon could be due to the habit of the curved neck, which slightly shift the resonance frequency with sound wave transformation under curvature. These finding suggested that the tapered neck set up with curved neck extension could reduce the resonance frequency without significant drop of absorption coefficient, however, no significant improvement of sound absorption performance been observed.

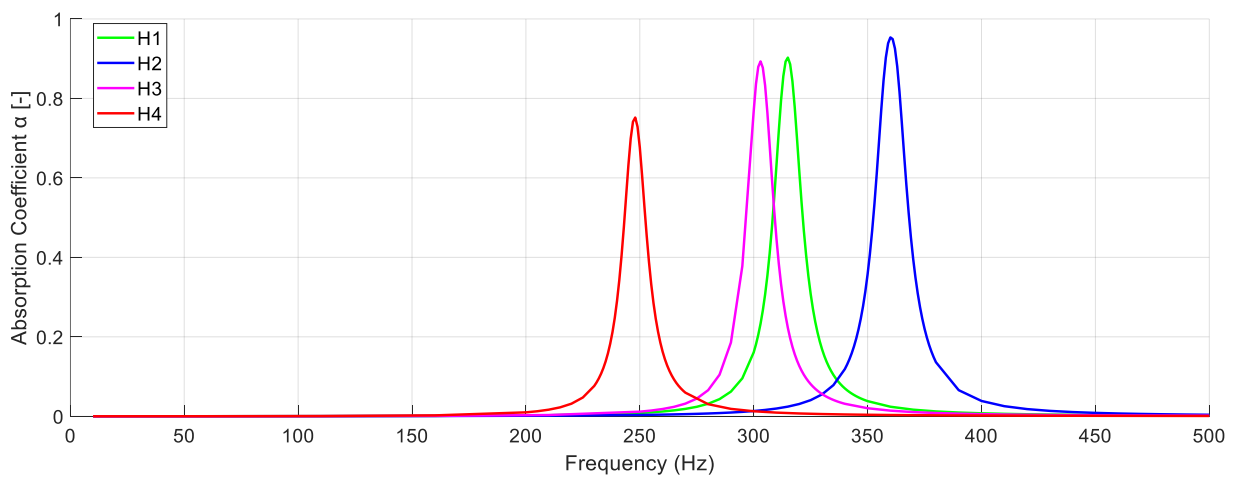

Figure 11: Plot of absorption coefficient of H1-H4 in the frequency range 10-500Hz.

Next step is to investigate the property of the tapered curved embedded neck resonator by carrying the parametric study with the COMSOL numerical analysis model. The ratio of outlet diameter to inlet diameter ($d_{n2}/d_{n1}$) was examined first, with other parameters remaining constant. As the ratio increases, the neck shape transitions from divergent ($d_{n2} < d_{n1}$) to non-tapered ($d_{n2} = d_{n1}$) and eventually to convergent ($d_{n2} > d_{n1}$), as illustrated by the a-axis in Figure 12. This analysis aims to determine the influence of the resonance frequency on $d_{n2}/d_{n1}$, specifically the changes in resonance frequency due to the divergent neck, non-tapered neck, and convergent neck from outlet to inlet. In this analysis, the value of $d_{n2}$ consist at 10mm, and the value of $d_{n1}$ changes to achieve the alternation of the ratio $d_{n2}/d_{n1}$.

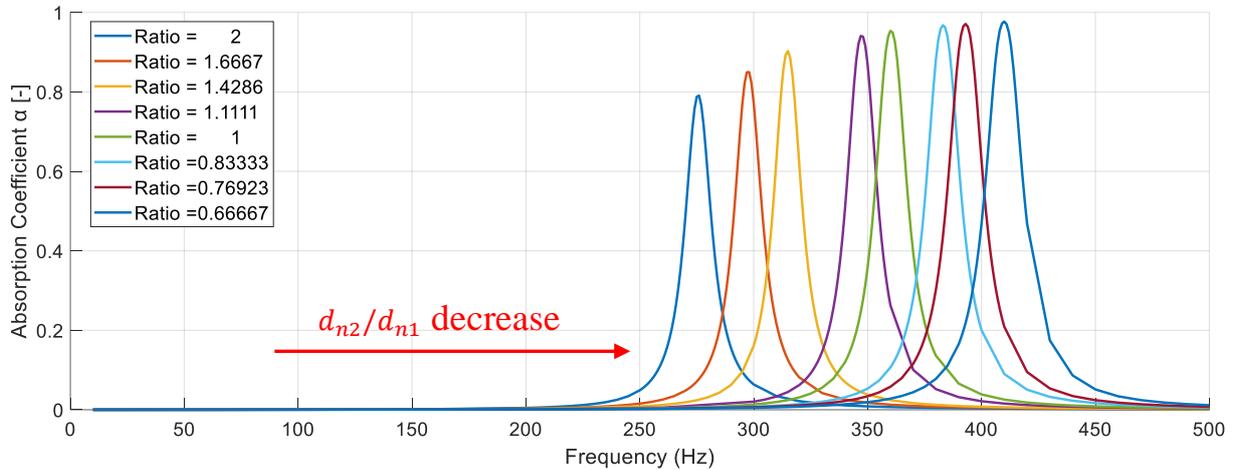

Figure 12: The plots of absorption coefficients as the ratio of $d_{n2}/d_{n1}$ decrease.

Figure 12 presents a plot of the absorption coefficient against frequency for different ratio parameters. Interpreting this graph, the resonance frequency and corresponding absorption coefficient both increase as the neck shape transitions from convergent to divergent. The general shape of the peaks does not have big difference, this concluded that the tapered geometry of curved neck does not enhance the bandwidth of the absorption ability. A quadratic regression analysis of the frequency-ratio relationship been carried with MATLAB and shown by Figure 13, which indicated the data fit well with the quadratic equation $y = 48.03x^2 - 213.37x + 531.64$ with R squared value of 0.9993. This indicated that there is quadratic relationship between the ratio and resonance frequency. However, as there is only 8 data points with smaller sample size during the analysis, the confidence in the relationship might be lower.

In summary, for this stage of the parametric analysis, the tapered curved embedded neck configuration of a Helmholtz resonator exhibits lower absorption frequency when the outlet/inlet ratio ($d_{n2}/d_{n1}$) is increased to attain a more convergent geometry. Conversely, decreasing the ratio to achieve a divergent geometry result in a higher absorption coefficient but also a higher resonance frequency.

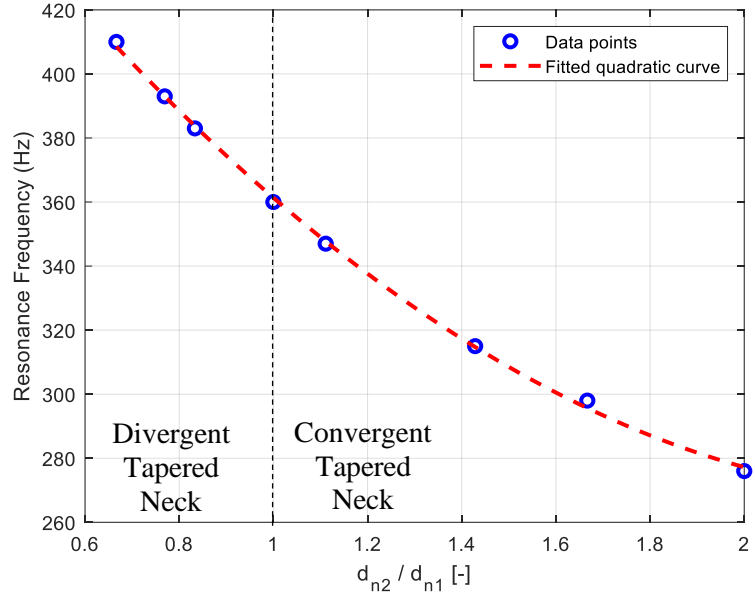

Figure 13: The corresponding resonance frequency for each ratio $d_{n2}/d_{n1}$ and quadratic regression analysis of the frequency-ratio relationship.

The second step of the parametric analysis is to investigate the relationship of changing effective neck length $l_{eff}$ to the resonance frequency. The effective length is the length of central line of the circular neck tube, Same procedure as described above, the other uncorrelated parameters keep constant except the value of effective length $l_{eff}$. The value of effective length varies from 22.366mm to 50.799mm, which equivalent to the 0.1-0.4 revolution of the curved neck plus the fundamental length due to the turning section of the neck. Figure 14 is the plot of the analysis with the 6 different length models. From observation, longer effective length leads to lower resonance frequency, but the maximum absorption coefficient is also reduced from 0.9426 to 0.8478 accordingly.

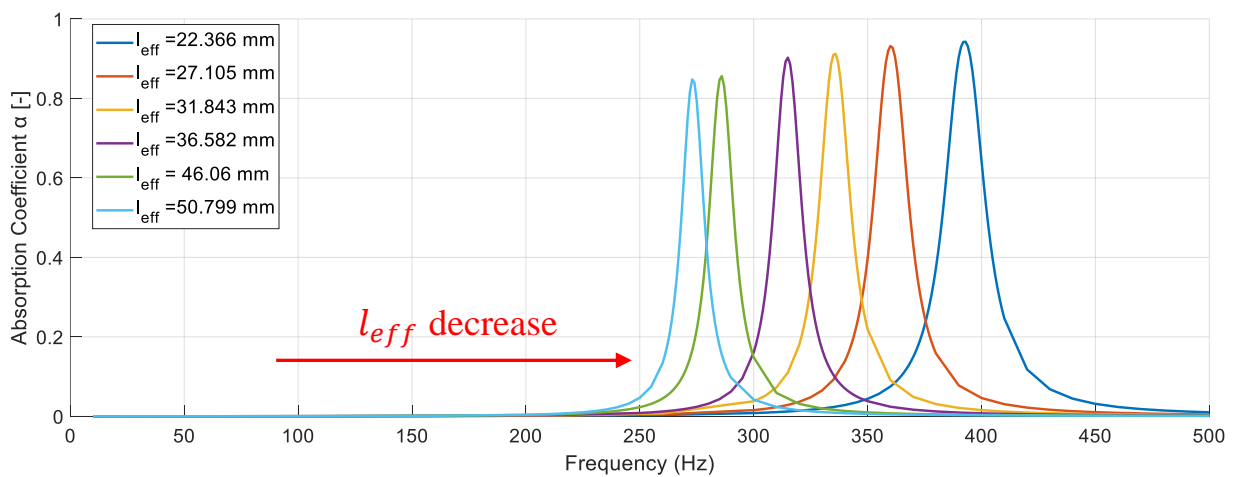

Figure 14: The plots of absorption coefficients as the effective length $l_{eff}$ decrease.

Previous literature [5] [13] states that, as the effective length increase, the resonance frequency decreases subsequently, and according to Equation (1), the length-frequency relationship should follow the quadratic relationship of $l_{eff} = a/(f^2 + c)$. In order to validate the properties of extending the neck length described by the literature and equation, a nonlinear least square fitting technique could be used to investigate the numerical relationship of the frequency corresponding to the effective length. The analysis been carried by MATLAB and shown by the plot in Figure 15. As the plot shows, the fitted curve been plotted according to the 8 data points, with quadratic expression $l_{eff} = 3209614.741/(f^2 - 11631.217)$. This expression is reasonable, as known from Equation (2), the constant $a$ is formed with squared speed of sound $c_0$ and ratio of squared diameter $d_n^2/d_0^2$, and constant c derived from the $\delta_n$ is the end correlation factor, which both match the significant size of the constant if $l_{eff}$ chose in millimetre. These groups of data have R squared value 0.995, which shows a good fitness of the data to the expression. In conclusion for this stage of parametric analysis, the tapered curved extended neck resonator shares the similar physical properties as the normal Helmholtz resonator configuration, which have significant reduction of resonance frequency as the neck length increase follow the quadratic relationship. However, as effective length increase, the absorption coefficient consequently reduced with a small quantity.

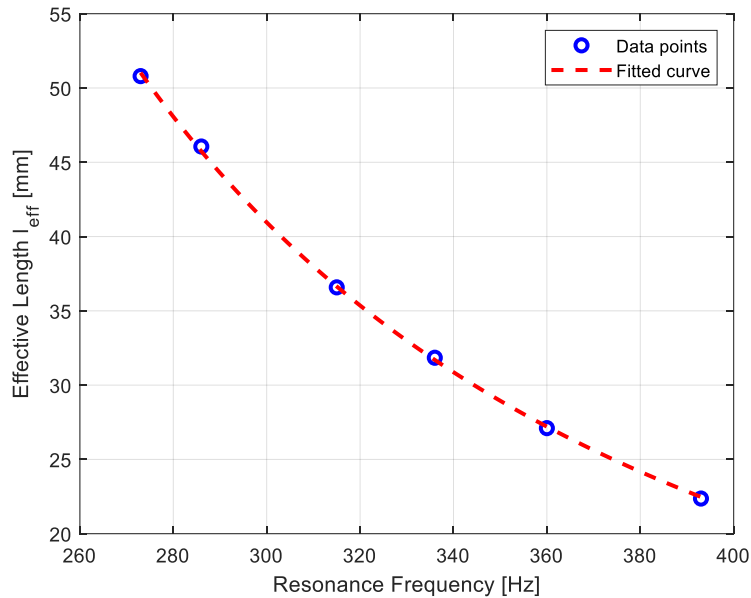

Figure 15: The nonlinear least fitting analysis for the frequency-length quadratic relationship.

The last stage of the parametric analysis is to investigate the general diameter of the neck under constant inlet/outlet ratio $d_{n2}/d_{n1}$ effect on the resonance frequency and absorption of the resonator. In this analysis, the $d_{n2}/d_{n1}$ ratio constraint at 10/7 and adjust the value $d_{n2}$ from 7mm to 10.5mm, with $d_{n1}$ changes with $d_{n2}$ dependently. Apply the same numerical analysis with COMSOL, the absorption-frequency graph been plotted by Figure 16 with 5 samples of different configurations. These plots indicated that as the general diameter of the embedded neck increases, the resonance frequency increase and absorption coefficient increase accordingly.

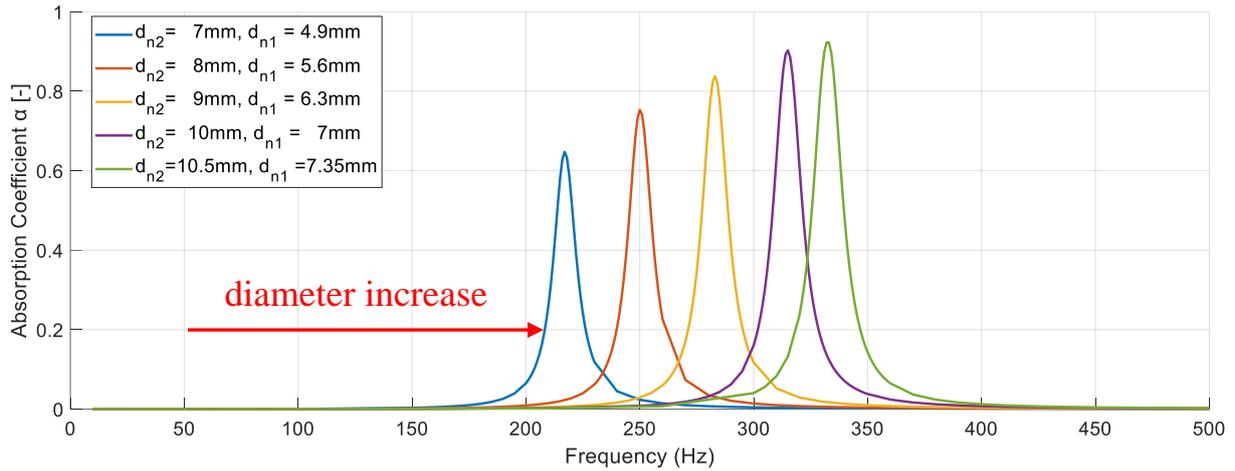

Figure 16: The plots of absorption coefficient as the diameter increase.

The derived Equation (2) for the fundamental configuration of the circular concentric Helmholtz resonator states that, the resonance frequency of the resonator is proportional to the neck diameter, which follow the linear relationship $f = md_n$. To prove this relationship, a Linear Regression Analysis been employed with MATLAB, and plot a best fit frequency-diameter line been plotted, with expression $f = 32.768d_n$, shown by Figure 17. The R square value of the plot is approximately to 1, which means the linear relationship between the resonance frequency and diameter of the neck is valid to apply in the tapered curved embedded neck configuration. In conclusion for the analysis of the diameter of neck, smaller diameter could reduce the resonance frequency linearly, however, absorption performance of the resonator will be reduced correspond to the frequency.

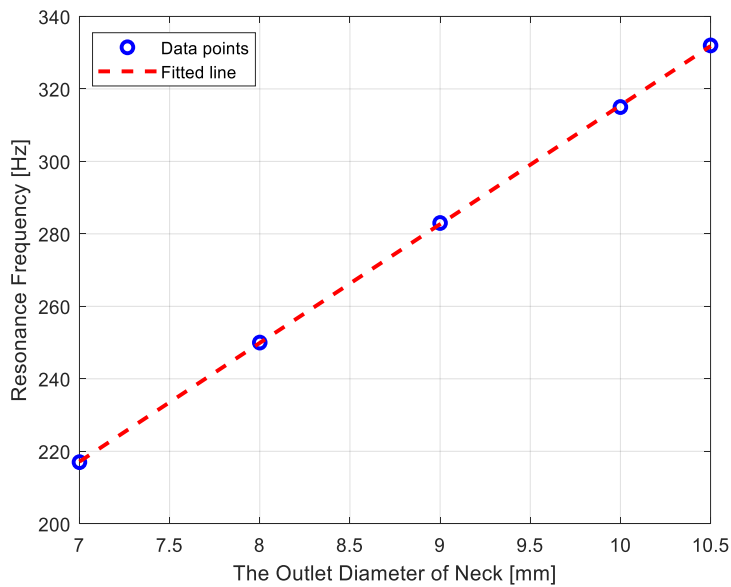

Figure 17: The linear regression analysis of the frequency-diameter relationship for the data.

## 3.2 Experimental Validation

Three representative models picked from the numerical parametric analysis, with a baseline model were validated by the two-microphone acoustic impedance tube. The parameter of the four 3D-printed model were shown by Table 3 above, with numerical and experimental results shown by the Figure 18.

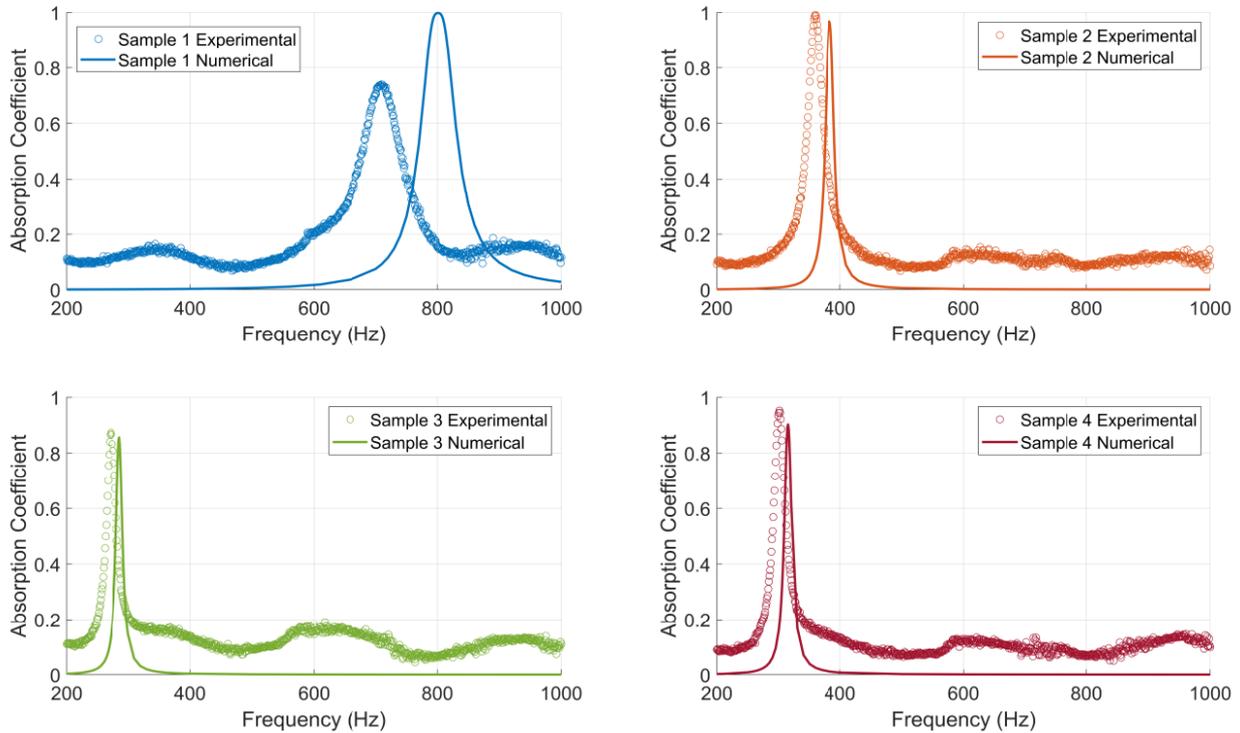

Figure 18: The experimental and numerical analysis of the sample 1-4.

In this experimental validation, sample 1 is the baseline model, while samples 2-4 are selected from the numerical parametric analysis. The values of interest are presented in Table 5, which compares experimental and analytical results. Generally, the resonance frequency from the experiment is slightly lower than the numerical results, and the maximum absorption coefficient for samples 2-4 is higher than the numerical findings apart from the baseline model. It is evident that the maximum absorption coefficient value and resonance frequency of the baseline model is not accurate due to the significant data difference. This is probably due to the manufacture error on the baseline sample. Read from the Figure 18, the experimental values follow the trend of the numerical analysis, which validate the trends described during the parametric study. However, the values do not perfectly align with the numerical plot, as there are minor shifts towards lower frequencies for the peak. This deviation may be due to the precision of the 3D printing process. Caused by the complexity of the geometry of the resonator, the neck and wall for the resonator may collapse during the printing process, and lead to the geometry difference to the CAD input. Moreover, due to the exclusive use of glue to seal the cavity with a lid, the thickness of the glue layer may lead the lid positioned higher than expected and result the peak shift to lower frequency.

When comparing the baseline (sample 1) to samples with tapered curved extended neck configurations (samples 2-4), a significant decrease in resonance frequency is observed. The comparison between samples 2 and 4 reveals that implementing a divergent tapered neck leads to a higher resonance frequency and increased resonance absorption, which is consistent with the findings from the parametric analysis. Additionally, comparing samples 3 and 4 demonstrates that the resonance frequency decreases as the effective length decreases. Therefore, based on the experimental results, the trends in the properties identified in the parametric analysis are confirmed to be valid.

Table 5: The results analysed by COMSOL and experiment, with percentage difference for each sample.

|  | Sample 1 | Sample 2 | Sample 3 | Sample 4 |
|---|---|---|---|---|
| Resonance Frequency | 801 | 383 | 286 | 315 |
|  | 710.25 | 359.25 | 272.25 | 300.75 |
|  | 12.01% | 6.40% | 4.93% | 4.63% |
| Absorption Coefficient | 0.9984 | 0.9678 | 0.8561 | 0.9027 |
|  | 0.7448 | 0.9919 | 0.8774 | 0.9546 |
|  | 29.10% | 2.46% | 2.46% | 5.59% |

## 4. CONCLUSION AND FUTURE WORKS

In this study, the performance of Helmholtz resonators with curved tapered neck extensions was investigated using both numerical and experimental approaches. The numerical parametric analysis was conducted using the FEM with COMSOL, and the results been validated through experimental tests using a two-microphone acoustic impedance tube. The parametric analysis examined the effects of the outlet diameter of the neck, the length of the neck, and the ratio on the absorption coefficients and resonance frequency of the resonators. The results indicated that the tapered curved embedded neck configuration exhibits lower absorption frequency when the outlet/inlet ratio is increased, and the neck geometry trend to convergent results in a lower resonance frequency. The analysis also shows that longer effective length of neck and smaller diameter of the neck could enhance the lower frequency noise absorption accordingly. The trends observed in the parametric analysis been confirmed with the experimental validation.

Overall, this study demonstrates the nature properties of the tapered curved embedded neck configuration of Helmholtz resonators, which provide a unique method of achieve lower resonance frequency without changing the size of cavity. This neck geometry has potential application on the wide-range sound absorption panels, which could achieve broadband absorption frequency range with different configurations of resonator units. Further research should focus on improving the fabrication process and conducting more extensive experimental validation to confirm the observed trends and explore potential applications for enhanced low-frequency noise attenuation capabilities. Also, the background theory needs to be derived for these types of resonators, which could provide a more comprehensive understanding and support a more solid validation for the conclusion.

# 5. ACKNOWLEDGEMENT

The author would like to acknowledge Professor Mahdi Azarpeyvand to provide valuable advises on this research, and thank for the PhD student Abhishek Gautam, who give a technical support on the experiment for this research.